\documentclass[iop]{emulateapj}

\def\lapp{\ifmmode\stackrel{<}{_{\sim}}\else$\stackrel{<}{_{\sim}}$\fi}
\def\gapp{\ifmmode\stackrel{>}{_{\sim}}\else$\stackrel{>}{_{\sim}}$\fi}

\newcommand{\source}{Swift~J1822.3$-$1606}
\newcommand{\src}{Swift~J1822.3$-$1606}
\newcommand{\rxte}{\textit{RXTE}}
\newcommand{\xte}{\textit{RXTE}}
\newcommand{\cxo}{\textit{Chandra}}
\newcommand{\chandra}{\textit{Chandra}}
\newcommand{\rosat}{\textit{ROSAT}}
\newcommand{\swift}{\textit{Swift}}
\newcommand{\xmm}{\textit{XMM-Newton}}
\newcommand{\tempo}{{\tt{TEMPO}}}
\newcommand{\degrees}{^{\circ}}

\newcommand{\lsk}{Paper~I}

\usepackage{amsmath}

\begin{document}

\title{Post-outburst X-ray flux and timing evolution of Swift~J1822.3$-$1606}

\author{
P. Scholz\altaffilmark{1},
C.-Y. Ng,
M. A. Livingstone,
V. M. Kaspi,
A. Cumming
\& R. Archibald
}
\affil{Department of Physics, Rutherford Physics Building,
McGill University, 3600 University Street, Montreal, Quebec,
H3A 2T8, Canada}

\altaffiltext{1}{pscholz@physics.mcgill.ca}

\begin{abstract}

\src\ was discovered on 2011 July 14 by the \swift\ Burst Alert Telescope following 
the detection of several bursts. The source was found to have a period of 8.4377\,s and
was identified as a magnetar. Here we present a phase-connected timing analysis
and the evolution of the flux and spectral properties using \xte, \swift, and
\chandra\ observations. We measure a spin frequency of 0.1185154343(8)\,s$^{-1}$ and
a frequency derivative of $-4.3\pm0.3\times10^{-15}$ at MJD 55761.0, in a timing analysis
that include significant non-zero second and third frequency derivatives 
that we attribute to timing noise. This corresponds
to an estimated spin-down inferred dipole magnetic field of
$B\sim5\times10^{13}$\,G, consistent with previous estimates though still possibly 
affected by unmodelled noise.
We find that the post-outburst 1--10\,keV flux evolution can be characterized by 
a double-exponential decay with decay timescales of $15.5\pm0.5$ and $177\pm14$ days.
We also fit the light curve with a crustal cooling model which suggests that the 
cooling results from heat injection into the outer crust.
We find that the hardness-flux correlation observed in magnetar outbursts also
characterizes the outburst of \src.
We compare the properties of \src\ with those of other magnetars and their
outbursts.

\end{abstract}


\section{Introduction}

Over the past two decades, several new classes of neutron stars have been
discovered \citep[see][for a review]{kas10}. Perhaps the most exotic is that
of the magnetars, which exhibit some highly unusual properties, often
including violent outbursts and high persistent X-ray luminosities that
exceed their spin-down powers \citep[for reviews see][]{wt06,mer08}. These
objects, while previously classified as anomalous X-ray pulsars (AXPs) and
soft gamma repeaters (SGRs), are now generally accepted as a unified class of
neutron stars powered by the decay of ultra-strong magnetic fields \citep[e.g.][]{tlk02}.

To date, there are roughly two dozen magnetars and candidates
observed,\footnote{See the magnetar catalog at
\url{http://www.physics.mcgill.ca/$\sim$pulsar/magnetar/main.html}.} with spin
periods between 2 and 12\,s, and high spin-down rates that generally suggest
dipole $B$-fields of order $10^{14}$ to $10^{15}$\,G \citep[except SGR
0418+5729;][]{ret+10}. Thanks to the \emph{Swift} satellite, several new magnetars
have been discovered in recent years via their outbursts
\citep[e.g.][]{rit+09,gcl+10,kkp+12}. Once a new source has been identified,
long-term monitoring is crucial to measure its timing properties, and hence to
constrain the dipole magnetic field strength. Also, the flux evolution
following an outburst could provide insights into many physical properties,
such as the location of energy deposition during an outburst, the crust
thickness and heat capacity \citep[see][Cumming et al. in prep.]{pr12},
or the physics of a highly active magnetosphere \citep{bel09,pbh12}.

One of the latest additions to the list of magnetars is \src. This source was
first detected by \emph{Swift} Burst Alert Telescope (BAT) on 2011 July 14
(MJD~55756) via its bursting activities \citep{cbc+11}. 
It was soon identified as a new magnetar 
upon the detection of a pulse period $P$=8.4377\,s \citep{gks11}. 
No optical counterpart was found, with 3$\sigma$ limit down to
a z-band magnitude of 22.2 \citep{rmie11}.  In \citet[][hereafter \lsk]{lsk+11}, we
reported initial timing and spectroscopic results using follow-up X-ray
observations from \swift, \emph{Rossi X-ray Timing Explorer} (\emph{\rxte}),
and \emph{Chandra X-ray Observatory}. We found a spin-down rate of $\dot
P=2.54\times 10^{-13}$ which implies a surface dipole magnetic
field\footnote{The surface dipolar component of the $B$-field can be estimated by
$B=3.2\times10^{19}(P\dot P)^{1/2}$\,G.} $B=4.7\times 10^{13}$\,G, the second
lowest $B$-field among magnetars.
Using an additional 6 months of \swift\ and \xmm\ data, \citet{rie+12} 
present a timing solution
and spectral analysis. They find a spin-down rate of $\dot P=8.3\times 10^{-14}$
which implies a magnetic field of $B=2.7\times 10^{13}$, slightly lower than 
that found in \lsk.


In this paper, we present an updated timing solution,
and the latest flux evolution using new observations from the same X-ray
instruments as in \lsk. The additional two \cxo\ and 18 \swift\ observations
provide a timing baseline that is over four times longer and allows a 
detailed study of the flux decay. We also report on an archival
\emph{ROSAT} observation to constrain the pre-outburst flux.
We discuss the effects of timing noise
on our timing solution and the properties and implications of this outburst 
within the magnetar model.

\section{Observations}

\begin{deluxetable*}{lccccc}
\tablecaption{Summary of observations of \src.
\label{ta:obs}}
\tablewidth{0pt}
\tabletypesize{\scriptsize}
\tablehead{\colhead{ObsID} & \colhead{Mode} & \colhead{Obs Date} & \colhead{MJD}   & \colhead{Exposure} & \colhead{Days since trigger}\\
           \colhead{}      & \colhead{}     & \colhead{}         & \colhead{(TDB)} & \colhead{(ks)}     & \colhead{} } 
\startdata
\hline
\multicolumn{6}{c}{\cxo}\\
\hline
{12612} & ASIS-S CC& 2011-07-27 &      55769.2 & 15.1 & 12.6\\
13511   & HRC-I    & 2011-07-28 &  55770.8 &  1.2 &  14.2\\ 
{12613} & ASIS-S CC& 2011-08-04 &    55777.1 & 13.5 & 20.5\\
{12614} & ASIS-S CC& 2011-09-18 & 55822.7 & 10.1 & 66.1\\
{12615} & ASIS-S CC& 2011-11-02 &  55867.1 & 16.3 & 110.5\\
{14330} & ASIS-S CC& 2012-04-19 &  56036.9 & 20.0 & 280.4\\
\hline
\multicolumn{6}{c}{\rosat}\\
\hline
{rp500311n00} & & 1993-09-12 & 49242 & 6.7 & --\\
\hline
\multicolumn{6}{c}{\swift}\\
\hline
00032033001 & PC & 2011-07-15 & 55757.7 & 1.6 & 1.2 \\
00032033002 & WT & 2011-07-16 & 55758.7 & 2.0 & 2.1 \\
00032033003 & WT & 2011-07-17 & 55759.7 & 2.0 & 3.1 \\
00032033005 & WT & 2011-07-19 & 55761.1 & 0.5 & 4.6 \\
00032033006 & WT & 2011-07-20 & 55762.0 & 1.8 & 5.5 \\
00032033007 & WT & 2011-07-21 & 55763.2 & 1.6 & 6.7 \\
00032033008 & WT & 2011-07-23 & 55765.8 & 2.2 & 9.2 \\
00032033009 & WT & 2011-07-24 & 55766.2 & 1.7 & 9.7 \\
00032033010 & WT & 2011-07-27 & 55769.5 & 2.1 & 12.9 \\
00032033011 & WT & 2011-07-28 & 55770.3 & 2.1 & 13.8 \\
00032033012 & WT & 2011-07-29 & 55771.2 & 2.1 & 14.7 \\
00032033013 & WT & 2011-07-30 & 55772.3 & 2.1 & 15.7 \\
00032051001 & WT & 2011-08-05 & 55778.0 & 1.7 & 21.5 \\
00032051002 & WT & 2011-08-06 & 55779.0 & 1.7 & 22.5 \\
00032051003 & WT & 2011-08-07 & 55780.4 & 2.3 & 23.9 \\
00032051004 & WT & 2011-08-08 & 55781.4 & 2.3 & 24.8 \\
00032051005 & WT & 2011-08-13 & 55786.4 & 2.2 & 29.8 \\
00032051006 & WT & 2011-08-14 & 55787.6 & 2.2 & 31.0 \\
00032051007 & WT & 2011-08-15 & 55788.1 & 2.3 & 31.6 \\
00032051008 & WT & 2011-08-16 & 55789.5 & 2.2 & 32.9 \\
00032051009 & WT & 2011-08-17 & 55790.3 & 2.2 & 33.8 \\
00032033015 & WT & 2011-08-27 & 55800.8 & 2.9 & 44.2 \\
00032033016 & WT & 2011-09-03 & 55807.2 & 2.4 & 50.6 \\
00032033017 & PC & 2011-09-18 & 55822.7 & 4.9 & 66.2 \\
00032033018 & WT & 2011-09-20 & 55824.5 & 1.5 & 68.0 \\
00032033019 & WT & 2011-09-25 & 55829.1 & 2.3 & 72.6 \\
00032033020 & WT & 2011-10-01 & 55835.1 & 2.6 & 78.5 \\
00032033021 & WT & 2011-10-07 & 55841.7 & 4.2 & 85.2 \\
00032033022 & WT & 2011-10-15 & 55849.2 & 3.4 & 92.7 \\
00032033023 & WT & 2011-10-22 & 55856.2 & 2.2 & 99.7 \\
00032033024 & PC & 2011-10-28 & 55862.2 & 10.2 & 105.6 \\
00032033025 & PC & 2012-02-19 & 55976.4 & 6.2 & 219.8 \\
00032033026 & WT & 2012-02-20 & 55977.0 & 10.2 & 220.5 \\
00032033027 & PC & 2012-02-21 & 55978.1 & 11.0 & 221.6 \\
00032033028 & WT & 2012-02-24 & 55981.9 & 6.7 & 225.4 \\
00032033029 & WT & 2012-02-25 & 55982.8 & 7.0 & 226.3 \\
00032033030 & WT & 2012-02-28 & 55985.0 & 7.0 & 228.5 \\
00032033031 & WT & 2012-03-05 & 55991.1 & 6.8 & 234.5 \\
00032033032 & WT & 2012-04-14 & 56031.1 & 4.3 & 274.6 \\ 
00032033033 & WT & 2012-05-05 & 56052.6 & 5.1 & 296.0 \\ 
00032033034 & WT & 2012-05-26 & 56073.0 & 4.9 & 316.5 \\ 
00032033035 & WT & 2012-06-17 & 56095.5 & 5.6 & 338.9 \\
00032033036 & WT & 2012-06-26 & 56104.1 & 6.2 & 347.6 \\
00032033037 & WT & 2012-07-06 & 56114.2 & 6.8 & 357.6 \\
00032033039 & WT & 2012-08-17 & 56156.1 & 4.9 & 399.6 \\
00032033040 & WT & 2012-08-22 & 56161.5 & 5.0 & 405.0 \\
\enddata
\end{deluxetable*}

\begin{deluxetable}{lcccc}
\tablecaption{Summary of \xte\ observations of \src.
\label{ta:rxteobs}}
\tablewidth{0pt}
\tabletypesize{\scriptsize}
\tablehead{\colhead{ObsID} & \colhead{Obs Date} & \colhead{MJD}   & \colhead{Exposure} & \colhead{Days since}\\
           \colhead{}      & \colhead{}         & \colhead{(TDB)} & \colhead{(ks)}     & \colhead{trigger} } 
\startdata
D96048-02-01-00& 2011-07-16 & 55758.49 & 6.5 & 1.96 \\
D96048-02-01-05& 2011-07-18 & 55760.81 & 1.7 & 4.28 \\
D96048-02-01-01& 2011-07-19 & 55761.57 & 5.1 & 5.04 \\
D96048-02-01-02& 2011-07-20 & 55762.48 & 4.9 & 5.95\\
D96048-02-01-04& 2011-07-21 & 55763.42 & 3.3 & 6.89\\
D96048-02-01-03& 2011-07-21 & 55763.64 & 6.0 & 7.11\\
D96048-02-02-00& 2011-07-22 & 55764.62 & 6.1 & 8.09\\
D96048-02-02-01& 2011-07-23 & 55765.47 & 6.8 & 9.94\\
D96048-02-02-02& 2011-07-25 & 55767.60 & 3.0 & 11.07\\
D96048-02-03-00& 2011-07-29 & 55771.35 & 6.8 & 14.82\\
D96048-02-03-01& 2011-08-01 & 55774.35 & 6.9 & 17.82\\
D96048-02-03-02& 2011-08-04 & 55777.85 & 1.9 & 21.32\\
D96048-02-03-04& 2011-08-04 & 55777.92 & 1.8 & 21.39\\
D96048-02-04-00& 2011-08-07 & 55780.49 & 6.9 & 23.96\\
D96048-02-04-01& 2011-08-09 & 55782.58 & 6.5 & 26.05\\
D96048-02-04-02& 2011-08-11 & 55784.97 & 3.7 & 28.44\\
D96048-02-05-02& 2011-08-12 & 55785.03 & 3.3 & 28.50\\
D96048-02-05-00& 2011-08-15 & 55788.05 & 5.9 & 31.52\\
D96048-02-05-01& 2011-08-16 & 55789.96 & 6.0 & 33.43\\
D96048-02-06-00& 2011-08-21 & 55794.46 & 6.6 & 37.93 \\
D96048-02-07-00& 2011-08-26 & 55799.61 & 6.8 & 43.1 \\
D96048-02-08-00& 2011-09-06 & 55810.38 & 6.0 & 53.8 \\
D96048-02-10-00& 2011-09-16 & 55820.24 & 6.7 & 63.7 \\
D96048-02-10-01& 2011-09-22 & 55826.18 & 5.6 & 69.6 \\
D96048-02-09-00& 2011-09-25 & 55829.38 & 6.2 & 72.8 \\
D96048-02-11-00& 2011-10-01 & 55835.90 & 7.1 & 79.4 \\
D96048-02-12-00& 2011-10-08 & 55842.23 & 5.9 & 85.7 \\
D96048-02-13-00& 2011-10-15 & 55849.67 & 5.6 & 93.1 \\
D96048-02-14-00& 2011-10-29 & 55863.11 & 6.7 & 106.6 \\
D96048-02-16-00& 2011-11-13 & 55878.90 & 5.9 & 122.4 \\
D96048-02-17-00& 2011-11-20 & 55885.21 & 6.0 & 128.7 \\
D96048-02-15-00& 2011-11-28 & 55893.18 & 6.7 & 136.6 \\
\enddata
\end{deluxetable}

\subsection{\swift\ Observations}

The \swift\ X-Ray Telescope (XRT) consists of a Wolter-I telescope and an {\em XMM-Newton} EPIC-MOS CCD detector \citep{bhn+05}.  
\swift\ is optimized to provide rapid follow-up of gamma-ray bursts and other X-ray transients.
Following the 2011 July 14 outburst of \src, the XRT was used to obtain 
46 observations for a total exposure time of 175\,ks.
Data were collected in two different modes, Photon Counting (PC) and Windowed
Timing (WT). While the former gives full imaging capability with a time resolution of 2.5\,s, the latter forgoes imaging
to provide 1.76-ms time resolution by reading out events in a collapsed 
one-dimensional strip.

For each observation, the unfiltered Level 1 data were downloaded from the \swift\ quicklook 
archive\footnote{http://swift.gsfc.nasa.gov/cgi-bin/sdc/ql}. 
For a summary of observations used, see Table \ref{ta:obs}.
The standard XRT data reduction script, {\ttfamily xrtpipeline}, was then run
using HEASOFT 6.11 and the \swift\ 20110725 CALDB.
We reduced the events to the barycenter using the position of
RA$=18^{\rm{h}}$~$22^{\rm{m}}$\,$18^{\rm{s}}$, 
Dec$=-16\degrees$\,$04\arcmin$\,$26\farcs8$ \citep{pbk11}.
Source and background events were extracted using the following regions:
for WT mode, a 40-pixel long strip centered on the source was used to extract the
source events and a strip of the same size positioned away from the source was
used to extract the background events.
For PC mode, a circular region with radius 20 pixels was used for the source
region and an annulus with inner radius 40 pixels and outer radius 60 pixels was used 
as the background region. For the first PC mode observation (00032033001), a circular
region with radius 6 pixels was excluded to avoid pileup.
For the subsequent PC mode observation (00032033017),
a region with radius 2 pixels was excluded. We estimate the maximum pileup fraction of the remaining
PC observations be less than 5\%.

For the spectral analysis, \swift\ ancillary response files, which provide the effective
area as a function of energy, were created using the 
FTOOL {\tt xrtmkarf} and the spectral redistribution matrices from the \swift\ CALDB 
were used.

\subsection{\rxte\ observations} 
\label{sec:rxteobs}

The \rxte\ Proportional Counter Array (PCA) comprised five proportional counting units 
and provided a large collecting area and high timing precision \citep{jmr+06}.
We downloaded 32 observations from the {\em HEASARC} archive spanning an MJD range from 55758 to 55893,
for a total of 174\,ks of integration time.
The data were collected in {\tt GoodXenon} mode which records each event with
1-$\mu$s time resolution. The observations are summarized in Table \ref{ta:rxteobs}. 

We selected events in the 2--10 keV energy range (PCA channels 6--14) from the top xenon layer of each PCU 
for our analysis, to maximize signal-to-noise ratio. 
The data from all the active PCUs were then merged. If more than one
observation occurred in a 24-hr period, the observations were combined into a single data set.
Photon arrival times were adjusted to the solar system barycenter using the same position
as the for \swift\ data. Events were then binned into time series with resolution 1/32 s for use
in the following analysis.

\subsection{\cxo\ observations}
\label{sec:cxoobs}

Following the outburst, we triggered our ToO program with the \emph{Chandra
X-ray Observatory}. The telescope onboard \emph{Chandra} has an effective area
$\sim$3 times larger than that of \swift\ XRT, when used with the ACIS detector
in continuous clocking (CC) mode. This mode has a time resolution of 2.85\,ms
and sensitivity between 0.3 and 10\,keV\footnote{\url{http://cxc.harvard.edu/proposer/POG/html/}}.
Five ACIS CC-mode observations were obtained between days 13 and 281 after the
outburst, with exposures ranging from 10 to 20\,ks. The observation parameters 
are summarized in Table~\ref{ta:obs}.
For imaging purposes, we also processed a short (1.2\,ks) archival \emph{Chandra}
HRC-I observation taken 14 days after the outburst.

All \emph{Chandra} data were processed using CIAO 4.3 with CALDB 4.4.6.  We
extracted the source events with a 6\arcsec-long strip region, 
and the remainder of the collapsed strip ($\sim$7\arcmin\ long), 
excluding the region within 1\arcmin\ of the source
in order to minimize any contamination from the wings of the PSF, 
was used for the background.
We restricted the timing analysis to events between 0.3 and
8\,keV. Photon arrival times were corrected to the solar system
barycenter. The source spectrum
was extracted using the tool \texttt{specextract}.

\subsection{\rosat\ observation}
The only existing X-ray image that covers the field prior to the outburst is
a 6.5-ks \emph{ROSAT} PSPC \citep{asc85} observation of the nearby H{\sc ii} region M17
(Omega Nebula, G15.1$-$0.8).
The observation has a time resolution 
of 130\,ms. We downloaded the filtered event list from the HEASARC
data archive\footnote{\url{http://heasarc.gsfc.nasa.gov/W3Browse/}} and
carried out the analysis using FTOOLS.

\section{Analysis \& Results}

\subsection{Imaging}

\begin{figure*}
\plottwo{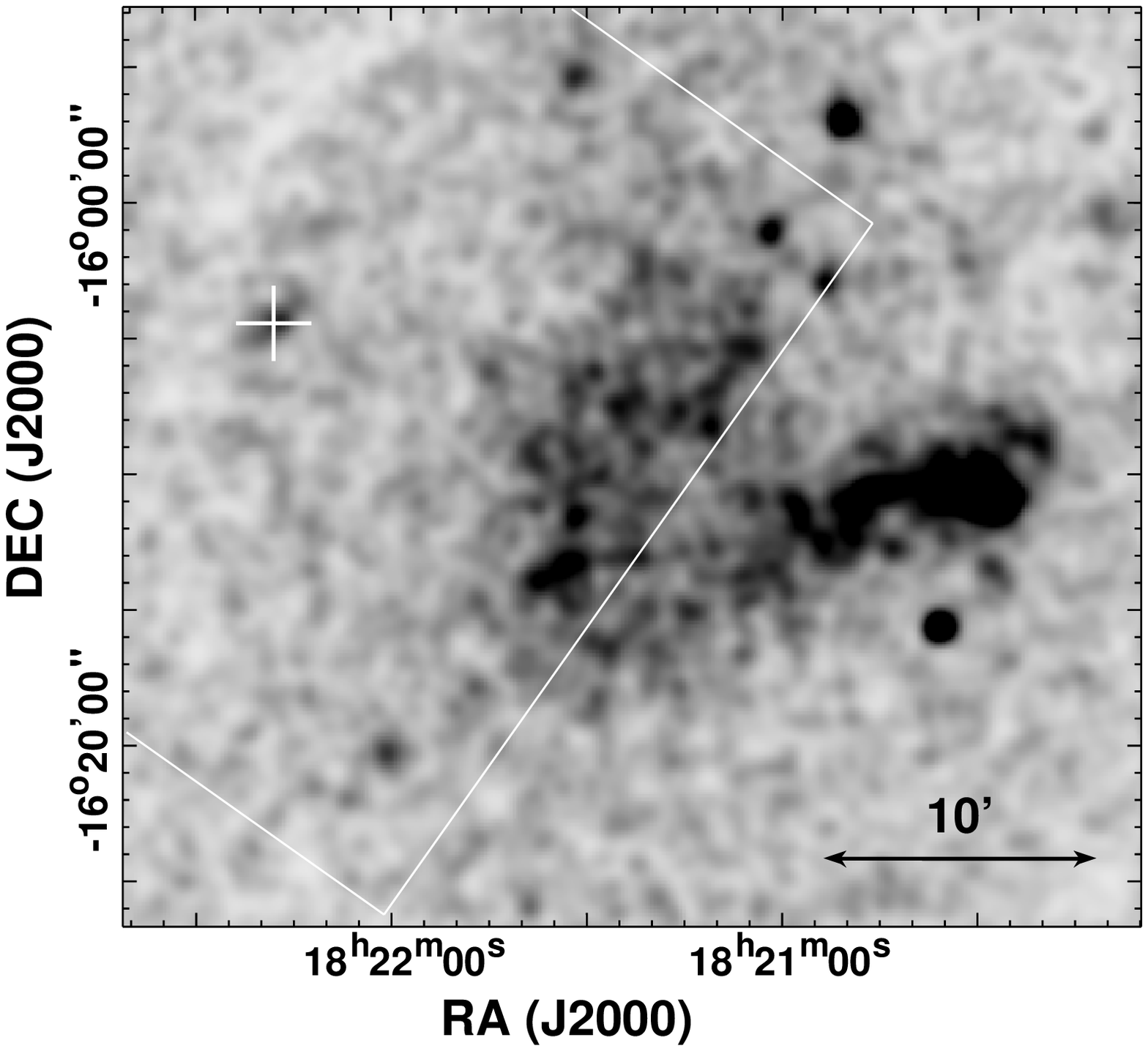}{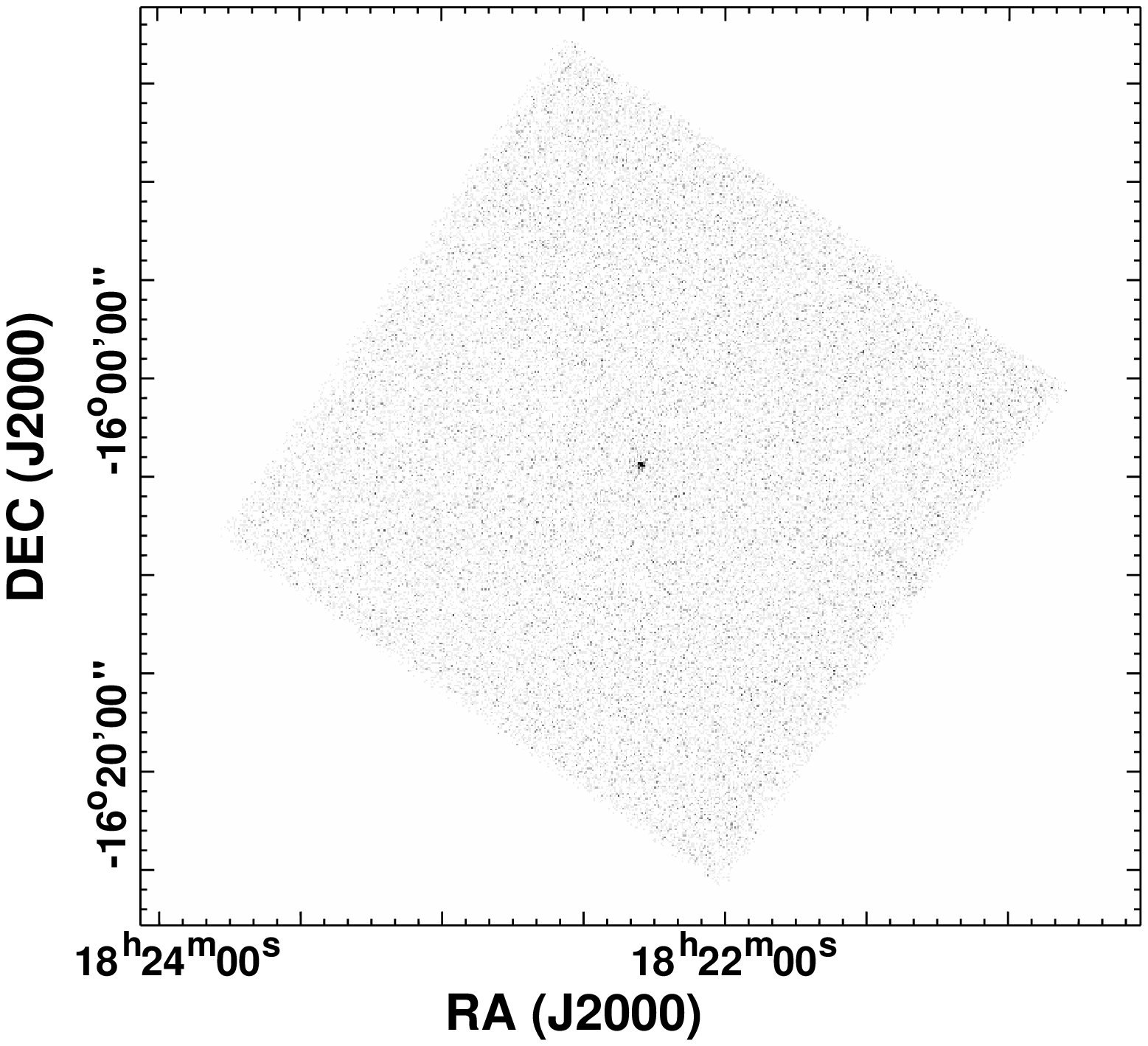}
\figcaption{
Left: \emph{ROSAT} image of the field of \src\ in 0.1--2.4\,keV
range. The position of \src\ is marked by the cross, and the lines indicate
the \emph{Chandra} HRC observation field of view. The large-scale diffuse
emission is the Galactic H{\sc ii} region M17.
Right: 1.2\,ks \emph{Chandra} HRC exposure of \src.
\src\ is the only source detected.
\label{fig:rosat}
}
\end{figure*}

Figure \ref{fig:rosat} shows the \emph{ROSAT} and \emph{Chandra} images. \src\ is the only
source detected in the \emph{Chandra} HRC image and its radial profile is
fully consistent with that of a model PSF. Hence, there is no evidence for any surrounding
nebula or dust scattering halo. We find a source position of
RA$=18^{\rm{h}}$~$22^{\rm{m}}$\,$18.06^{\rm{s}}$, 
Dec$=-16\degrees$\,$04\arcmin$\,$25\farcs55$
from the HRC image which is consistent
with the XRT position from \citet{pbk11} used above. 
We assume an error radius of $0.6\arcsec$ which is the uncertainty in the
absolute astrometry of \chandra\ for a 90\% confidence interval
\footnote{According to \url{http://cxc.harvard.edu/cal/ASPECT/celmon/}}.
In the \emph{ROSAT} image, an unresolved source is clearly detected at the position
of the magnetar, as first reported by \citet{erit11}. 
Since the \cxo\ image shows no other bright X-ray sources in the field, we take
this source to be \src\ in quiescence. Using a $4\arcmin \times 2\arcmin$ elliptical
aperture, we obtain $113\pm11$ total counts in 0.1--2.4\,keV range, of which $48\pm7$ counts
are due to background. Finally, we note that the diffuse X-ray emission
$\sim$20\arcmin\ southwest of the magnetar is from M17,
which contains the young stellar cluster NGC~6618 with over 100 OB stars
\citep{ldmg91}.

\subsection{Timing Analysis}
\label{sec:timing}

\subsubsection{Spin Evolution}
Barycentered events were used to derive a pulse time-of-arrival (TOA) for each \swift\ WT mode,
\rxte, and \cxo\ observation. For the \rxte\ observations, events were binned into
time series with 31.25-ms resolution. The time series were then folded with 128 phase
bins using the ephemeris from \lsk. A TOA was then measured
from each profile by cross-correlation with a template profile.
We verified that the \rxte\ pulse profiles were consistent with each other except in
one isolated observation which was handled accordingly (see Section \ref{sec:profiles}).

For \swift\ and \cxo\ observations, TOAs were extracted using a Maximum Likelihood (ML)
method, as it yields more accurate TOAs than the
traditional cross-correlation technique \citep[see][]{lrc+09}. 
This method was not used for the \rxte\
observations as their high number of counts (due to the large collecting area and background
count rates of the PCA) make the ML method computationally expensive.
The ML method for measuring TOAs requires a continuous model
of the template pulse profile for which we used a Fourier model.
The discrete Fourier Transform of the binned template profile was first calculated. 
The template was then fitted by $f(\phi)=\sum_{j=0}^n\alpha_je^{i 2\pi  j\phi}$
 where $\alpha_j$ is the Fourier coefficient for the $j^{th}$ harmonic, 
and $\phi$ the phase between 0 and 1. 
The number of harmonics used was optimized to account for the features of the light curve 
while ignoring small fluctuations caused by the finite number of counts. For
 \src, we used five harmonics to derive the TOAs. 

For each observation, a probability or likelihood
for a grid of trial offsets, $\phi_\mathrm{off}$, can be calculated using
$P(\phi_\mathrm{off}) = \prod^N_{i=0} f(\phi_i - \phi_\mathrm{off})$, where $\phi_i$ is
the phase of each photon folded at the best ephemeris of the pulsar. The likelihood
distribution that results then describes the probability density for the average pulse
arrival time. A TOA can be calculated from the optimal phase offset. 
We estimated TOA uncertainties by simulating one-hundred sets of events
drawn from the pulse profile of the observation and measured an offset
for each set using the ML method. 
The standard deviation of the simulated offset distribution
was then taken as the TOA uncertainty.
The ML derived TOAs were consistent with those derived for \lsk\ 
using the cross-correlation method. 

\begin{figure*}
\plotone{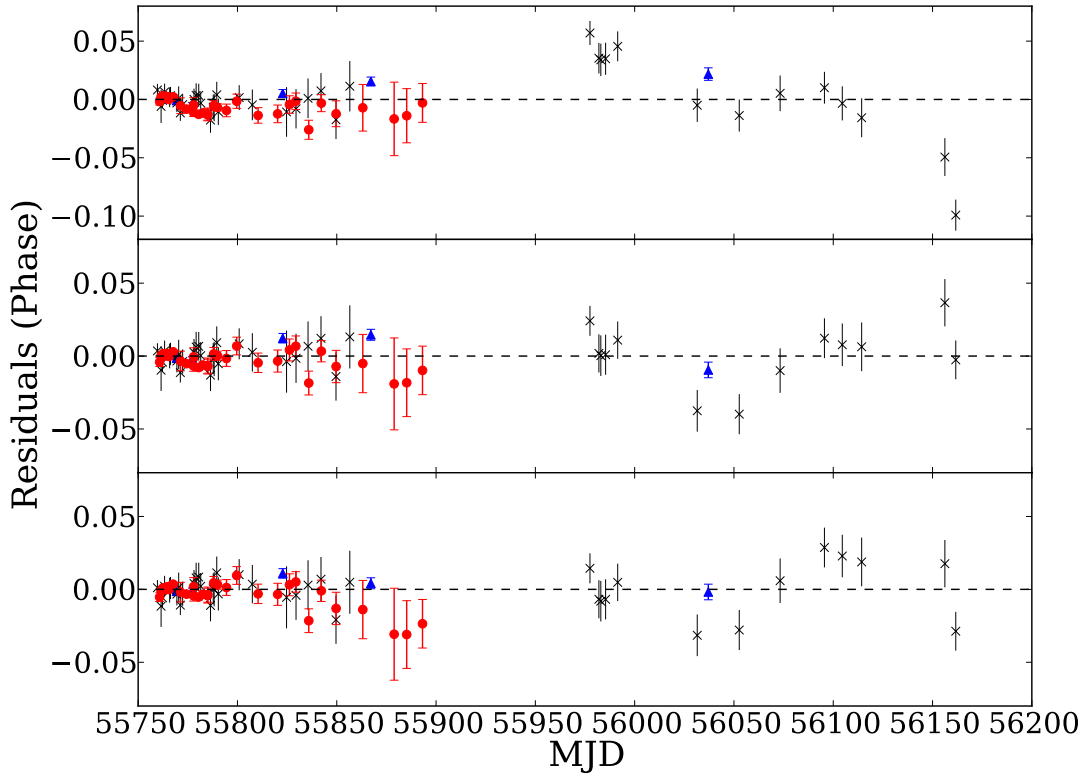}
\figcaption{
Timing residuals of \src. The top panel shows the residuals for the timing solution with one
frequency derivative (Solution 1); see Table \ref{ta:coherent}. 
The middle panel shows the residuals with the addition of a second
derivative (Solution 2). The bottom panel shows the residuals with three
frequency derivatives fitted. 
\swift\ WT mode data are represented by crosses, red circles 
denote \rxte\ observations and \cxo\ observations are shown as blue triangles.
\label{fig:resids}
}
\end{figure*}

Timing solutions were then fit to the TOAs using \tempo
\footnote{http://www.atnf.csiro.au/people/pulsar/tempo/}.
Three solutions, one with a single freqency derivative, one with two frequency derivatives
 and one with three frequency derivatives, 
are given in Table \ref{ta:coherent}. The top panel
of Figure \ref{fig:resids} shows the timing residuals with just $\nu$ and $\dot\nu$ 
fitted (Solution 1), the middle panel shows the residuals with $\ddot\nu$ also 
fitted (Solution 2), and the bottom panel shows the residuals with $\ddot\nu$ 
and $\dddot\nu$ also fitted (Solution 3). 
Solution 1 is a poor fit with a $\chi^2_\nu/\nu$ of 5.02/72.
This is likely due to timing noise, a common phenomenon in young neutron
stars including magnetars \citep[e.g.][]{dkg08, lk11}. 
The best-fit $\nu$ and $\dot\nu$ values for Solution 1 imply a 
surface dipolar magnetic field
of $B = 2.43\pm0.03 \times 10^{13}$\,G. Solution 2, with a significant non-zero $\ddot\nu$,
gives a better fit with a 
$\chi^2_\nu/\nu$ of 1.94/71.
An $F$-test gives a probability of $2 \times 10^{-16}$ that 
the addition of a second derivative 
does not significantly improve the fit. The surface dipolar magnetic field implied by
Solution 2 is $B = 3.84\pm0.08 \times 10^{13}$\,G. Solution 3, with a significant non-zero $\dddot\nu$,
provides still a better fit than Solution 2 with a $\chi^2_\nu/\nu$ of 1.44/70.
An $F$-test gives a probability of 
$3 \times 10^{-6}$ that the addition of a third derivative
does not significantly improve the fit. The best-fit parameters from Solution 3 imply a 
surface dipolar magnetic field of $B = 5.1\pm0.2 \times 10^{13}$\,G.

Note that the fit is heavily influenced by the very high quality \cxo\ TOAs.
However, omitting them and including only TOAs from \swift\ and \rxte\ still yields
significant second and third derivatives and an implied $B$-field of 
$B = 4.8\pm0.2 \times 10^{13}$\,G which is consistent with that of Solution 3.
The above-quoted uncertainties in $B$ 
and other derived quantities in Table \ref{ta:coherent} reflect only the 
statistical uncertainties in $\nu$ and its derivatives and do not include any
contributions from the simplified assumptions in the standard formulae used
to determine such quantities. Note that even with the addition of highly 
significant second and third derivatives, Solution 3 still does not provide an adequate fit.
Adding additional derivatives reduces the $\chi^2$ with marginal significance and
results in larger values of the spin-down rate and hence $B$. For example, including
a fourth frequency derivative does not result in significant improvement in $\chi^2$ 
($\chi^2_\nu/\nu = 1.31/69$) and yields $B=6.0\times10^{13}$\,G. 

To search for pulsations in the \rosat\ observation,
we applied a barycenter correction to the event arrival times, then
used the $Z_m^2$ test \citep{bbb+83} to search for pulsations.
We searched in the frequency range from zero to 3.8\,kHz in steps of 1.3\,$\mu$Hz,
oversampling the independent Fourier spacing by a factor of 10;
however, we found no significant signal. By simulating a pulsar with a
background subtracted count rate of that of the \rosat\ observation,
we find that the pulsar would be undetectable even with a pulsed fraction
of 100\%, therefore we cannot constrain the pulsed fraction.

\begin{deluxetable}{lc}
\tablecaption{Spin Parameters for \src.
\label{ta:coherent}}
\tablewidth{0pt}
\tabletypesize{\scriptsize}
\tablehead{\colhead{Parameter} & \colhead{Value}}
\startdata
Dates (Modified Julian Day)         & 55759~--~56161 \\
Epoch (Modified Julian Day)         & 55761.0 \\
Number of TOAs - \rxte              & 31 \\
Number of TOAs - \swift             & 40 \\
Number of TOAs - \cxo               & 5 \\
\cutinhead{Solution 1 - one frequency derivative}
$\nu$ (s$^{-1}$)                      &  0.1185154253(3)\\
$\dot{\nu}$ (s$^{-2}$)               & $-9.6(3)\times10^{-16}$ \\
RMS residuals (ms)                 & 52.2 \\
$\chi^2_\nu/\nu$                   & 5.02/72 \\
$B$ (G)                              & $2.43(3)\times10^{13}$ \\
$\dot{E}$ (erg s$^{-1}$)             & $4.5(1)\times10^{30}$\\
$\tau _c$ (kyr)                   & 1963(51)\\
\cutinhead{Solution 2 - two frequency derivatives}
$\nu$ (s$^{-1}$)                      & 0.1185154306(5) \\
$\dot{\nu}$ (s$^{-2}$)               & $-2.4(1)\times10^{-15}$ \\
$\ddot{\nu}$ (s$^{-3}$)               & $1.12(8)\times10^{-22}$ \\
RMS residuals (ms)                 & 32.2 \\
$\chi^2_\nu/\nu$                   & 1.94/71 \\
$B$ (G)                              & $3.84(8)\times10^{13}$ \\
$\dot{E}$ (erg s$^{-1}$)             & $1.12(5)\times10^{31}$\\
$\tau _c$ (kyr)                   & 784(33)\\
\cutinhead{Solution 3 - three frequency derivatives}
$\nu$ (s$^{-1}$)                      & 0.1185154343(8) \\
$\dot{\nu}$ (s$^{-2}$)               & $-4.3(3)\times10^{-15}$ \\
$\ddot{\nu}$ (s$^{-3}$)               & $4.4(6)\times10^{-22}$ \\
$\dddot{\nu}$ (s$^{-4}$)               & $-2.2(4)\times10^{-29}$ \\
RMS residuals (ms)                 & 27.5 \\
$\chi^2_\nu/\nu$                   & 1.44/70 \\
$B$ (G)                              & $5.1(2)\times10^{13}$ \\
$\dot{E}$ (erg s$^{-1}$)             & $2.0(2)\times10^{31}$\\
$\tau _c$ (kyr)                   & 442(33)\\
\enddata
\tablecomments{Errors are formal 1$\sigma$ \tempo\ uncertainties.}
\end{deluxetable}

\subsubsection{Pulse Profile Analysis}
\label{sec:profiles}

\begin{figure}
\plotone{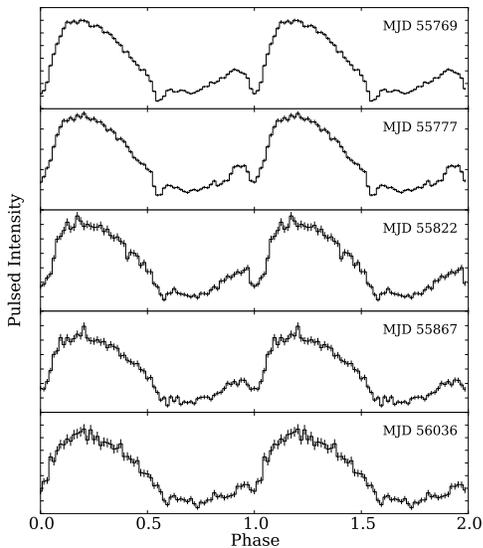}
\figcaption[]{Pulse profiles with 64 bins from each of the four \cxo\ observations.
The profiles are for the energy band 0.5--6\,keV. 
 \label{fig:chandra_profs}}
\end{figure}

\begin{figure}
\plotone{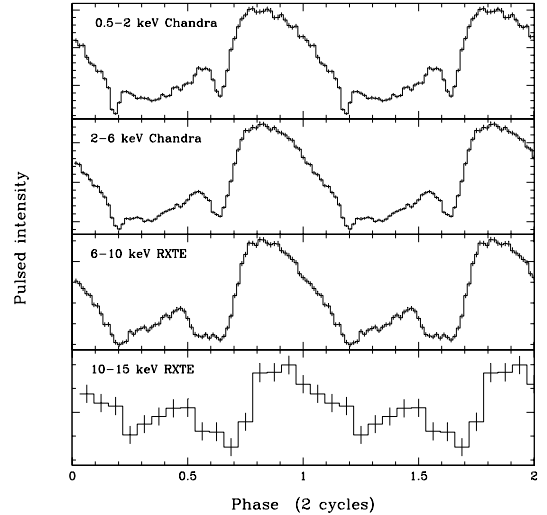}
\figcaption[]{Pulse profiles for \src\ from \cxo\ and \rxte\ data for
four energy ranges: 0.5--2\,keV and 2--6\,keV (\cxo\ data, 64 phase bins),
6--10\,keV and 10--15\,keV (\rxte\ data, 64 and 16 phase bins).
 \label{fig:energy_profs}}
\end{figure}

Here we search for time and energy
variability in the pulse profile of \src\ using the \rxte, \swift, and \cxo\ 
observations. 
We created pulse profiles for each
\rxte\ observation for energy ranges of 2~--~6\,keV, 6~--~10\,keV
 (with photons
selected from only the top xenon layer), 10~--~15\,keV, 15~--~20\,keV,
20~--~40\,keV, and 20~--~60\,keV (with photons selected from all
 three xenon layers) using the Solution 2 ephemeris.
For the \cxo\ data we produced pulse profiles with
the energy ranges of 0.5~--~6\,keV, 0.5~--~2\,keV and 2~--~6\,keV.
For the \swift\ data we created 0.5~--~10\,keV profiles, using
only WT mode observations as PC mode does not have sufficient time
resolution.

As in \lsk, we searched for time variability in pulse profiles but
found that all the \rxte\ profiles are consistent
with the template in each case except for the one profile from the 
very first observation after the outburst (Obsid D96048-02-01-00). 
The difference is due primarily to the off-pulse feature 
which had slightly different structure
between the template profile and the first \rxte\ observation.
We therefore did not use D96048-02-01-00 in the timing analysis.

The \chandra\ profiles, however, do show evidence for low-level variability.
Figure \ref{fig:chandra_profs} shows the 0.5~--~6\,keV pulse profile for each \cxo\ 
observation of \src. 
We produced residuals between each pair of \cxo\ profiles by
normalizing each profile and taking the difference between each normalized pair.
A comparison of the profile residuals between each set of profiles
shows that there is significant low-level evolution of the small `pulse' that precedes 
the main pulse. The main pulse does not exhibit any significant variation.
The most significant variability is that between the first (MJD~55769) and
last (MJD~56036) \cxo\ observation. The residuals between those two profiles have a 
${\chi}_\nu^2$ of 16.8 for 63 degrees of freedom. We note however that these
low-level variations in the smaller component likely do not have a significant impact
on the timing analysis, since the TOA extraction is heavily weighted toward the 
unchanging primary component. Indeed our simulations of the effects of such low-level
profile variations on the TOAs (see below) strongly support this conclusion. 

The \swift\ profiles also show evidence for low-level variability. As above,
we produced residuals between each pair of profiles and calculated a ${\chi}_\nu^2$
for the null hypothesis. The measured values are not consistent with a ${\chi}^2$
distribution, so there is significant variation between profile pairs. 
A closer look at the residuals shows that the variation is due primarily
to the small interpulse, as in the \cxo\ data.

To investigate the dependence of pulse morphology on energy, 
we created a single high significance
profile by aligning and summing individual profiles for each energy range.
Figure~\ref{fig:energy_profs} shows a summary of the results, with the
summed profiles for 0.5~--~2\,keV (top panel,
with 64 phase bins, \cxo\ data), 2~--~6\,keV (middle panel, with 64
phase bins, \cxo\ data), 6~--~10\,keV and 10~--~15\,keV (bottom two
 panels, \xte\ data,
with 64 and 16 phase bins, respectively).
No pulsations were detected above 15\,keV with the PCA.
We then calculated residuals between pairs of profiles, and
calculated ${\chi}_\nu^2$ values of the resulting residuals in 
order to identify energy dependence of the pulse morphology.
The most significant variability is between the 0.5~--~2\,keV \cxo\
profile and the 6~--~10\,keV \rxte\ profile. This can been seen in
Figure~\ref{fig:energy_profs} as a change in the phase of the
interpulse, arriving later for higher energies.  
For this profile pair, the
${\chi}_\nu^2$ of the residuals is 46.2 (for 28 degrees of
freedom), excluding the null hypothesis. The interpulse variability
causes significant differences between each pair of profiles,
except the 10~--~15\,keV profile, likely because of the lower statistics
of the latter.

\subsubsection{Comparison to Previously Reported Results}

\citet{rie+12} present a timing solution with a spin-down
implied magnetic field of $B=2.7\times10^{13}$\,G. Their data set is similar to ours,
although they use proprietary \xmm\ data whereas we use proprietary \chandra\ data 
and our data set includes seven additional \swift\ observations. Their timing solution 
is similar to our Solution 1. They, however, do not find a significant second frequency derivative.
A possible cause of this discrepancy could be the difference in TOA extraction methods.
Instead of using a pulse profile template, \citet{rie+12} fit the folded profile for
each observation with two sine functions with periods equal to the fundamental and the
first harmonic of the pulse period. They then assign the ascending node of the fundamental
sine function as the time-of-arrival of the pulse. 
This method was used to attempt to account for pulse-profile changes.
We implemented this method and derived
an additional set of TOAs to compare to our ML derived TOAs. We found that the
sine-model derived TOAs provided similar timing solutions as our ML TOAs and the addition
of a second frequency derivative did significantly improve the fit, reducing the $\chi^2_\nu/\nu$ 
from 7.91/72 to 2.72/71. The addition of a third derivative in this case, results in only
marginal improvement with a $\chi^2_\nu/\nu$ of 2.47/70.
If we limit our dataset to the \swift\ and \xte\ data used in \citet{rie+12}, we find
that the addition of a second frequency derivative is not necessary, which is consistent
with their findings.

In order to investigate the effects of pulse profile changes on both TOA extraction methods,
we simulated pulse profiles with an unchanging (other than noise) primary component and a
varying secondary component, as is observed in the pulse profile evolution of \src. We modelled
the profile using two gaussians and modified the amplitude of the smaller gaussian in order 
to vary the secondary component. We found that for both the sine-model and the ML methods,
as the amplitude of the secondary component was varied,
the measured phase offsets varied by less than their uncertainties. 
Hence we conclude that the observed pulse profile variations do not have an appreciable
affect on the TOA determination, independent of which TOA extraction method was used.

\subsection{Spectral Analysis \& Flux Evolution} 

\begin{deluxetable*}{lcccc}
\tablecaption{Models of the Flux Evolution of \src.
\label{ta:fluxmodel}}
\tablewidth{0pt}
\tabletypesize{\scriptsize}
\tablehead{\colhead{Model} & \colhead{$\tau_1$}\,(days) & \colhead{$\tau_2$}\,(days) & \colhead{$F_q$}\,(erg\,cm$^{-2}$\,s$^{-1}$) & \colhead{$\chi^2_\nu/\nu$}}
\startdata
Single Exponential & $23.8\pm0.5$ & -- & $3\times10^{-14}$ (fixed) & 20.1/43 \\ 
Single Exponential & $19.5\pm0.4$ & -- & $4.7\pm0.2\times10^{-12}$ & 4.48/42 \\ 
Double Exponential & $15.5\pm0.5$ & $177\pm14$ & $3\times10^{-14}$ (fixed) & 2.17/41 \\ 
Double Exponential & $9\pm1$ & $39\pm3$ & $4.0\pm0.2\times10^{-12}$ & 1.06/40 \\
\enddata
\end{deluxetable*}

\begin{figure}
\plotone{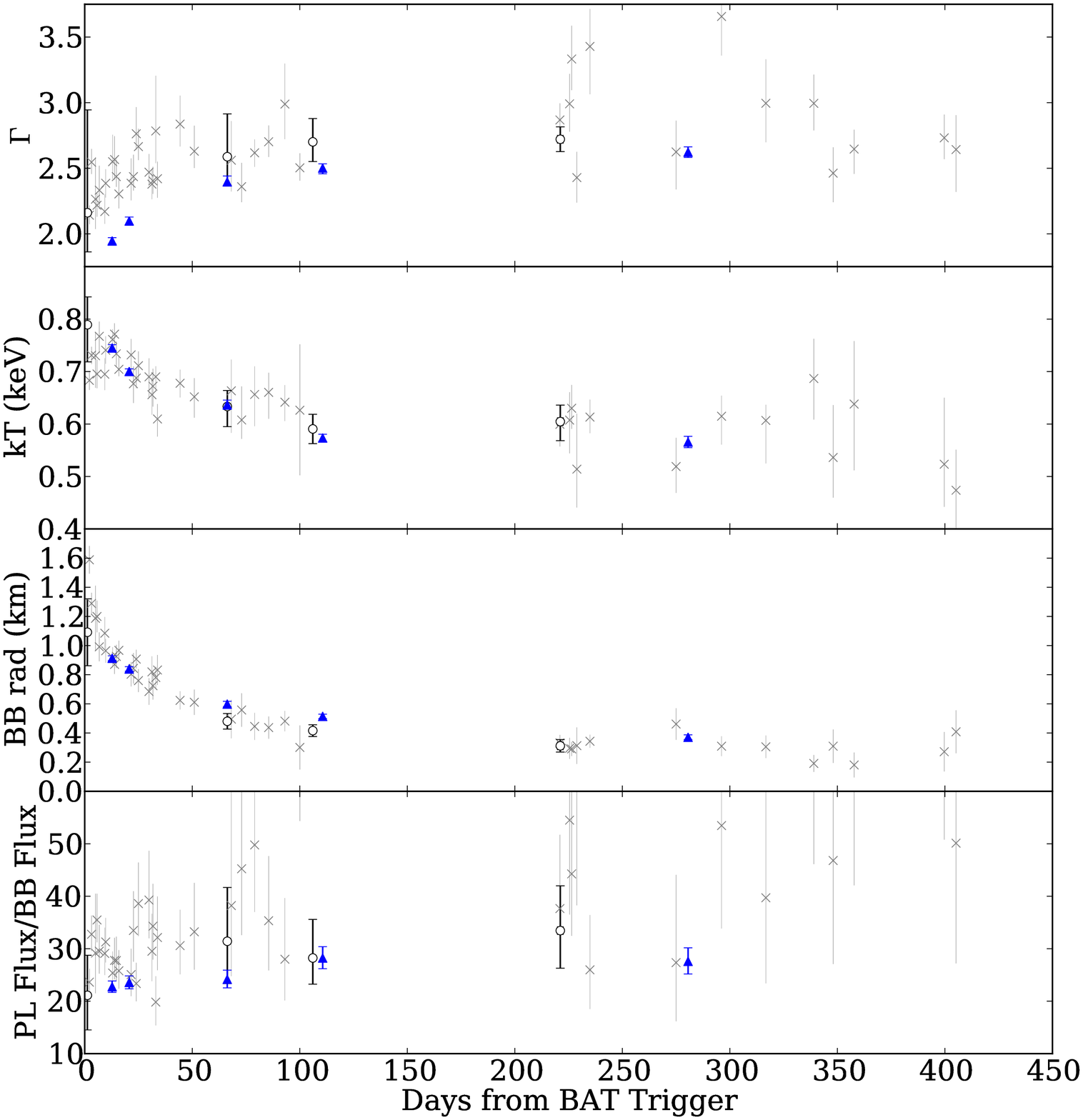}
\figcaption{
The spectral properties of \src\ following its outburst. Grey crosses represent \swift\ WT mode
data, open circles are \swift\ PC mode data, and blue triangles denote \cxo\ observations.
The determination of the blackbody radius assumes a distance of 1.6\,kpc (see Section \ref{sec:distance}). 
\label{fig:spec}
}
\end{figure}

\begin{figure}
\plotone{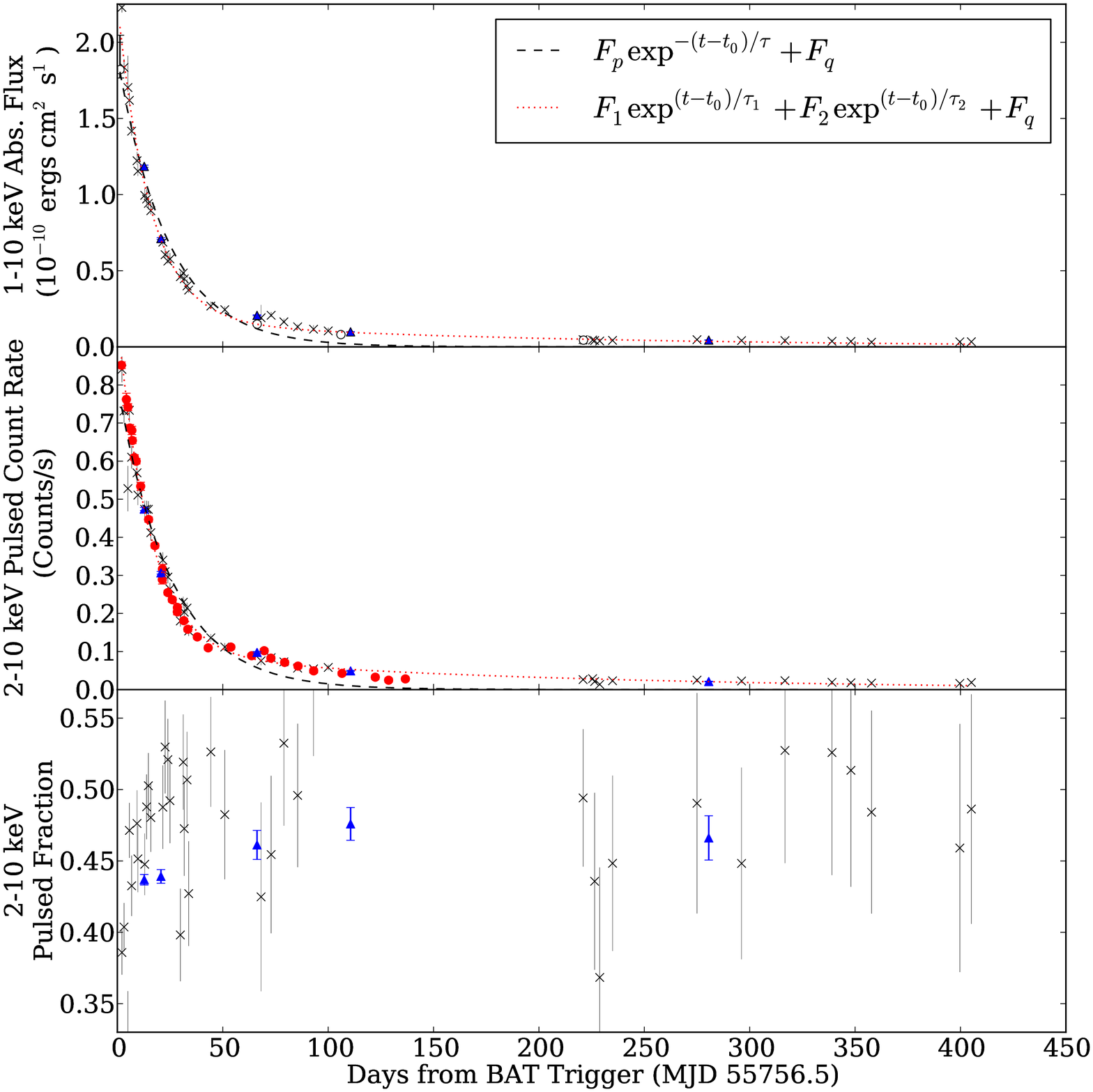}
\figcaption{
Flux evolution of \src\ following its outburst.
The top panel shows the 1--10\,keV absorbed flux as it decays. The middle panel
shows that 2--10\,keV pulsed count rate. The \rxte/PCA and \cxo\ count rates have been
scaled to the \swift\ count rates by fitting overall scaling factors
in the double-exponential fit. The bottom panel shows the pulsed fraction from
\swift\ and \cxo\ observations. In all panels,
\swift\ WT mode data are shown as grey crosses,
\swift\ PC mode data are represented by open circles,
\xte\ data are shown as red circles and data from \cxo\ are shown as blue triangles.
The dotted lines show the result of the double-exponential fit to the total and 
pulsed flux decays, and the dashed line shows the results of the single-exponential fit.
\label{fig:flux}
}
\end{figure}

\label{sec:spectral}

Spectral models were fit to the \swift, \cxo\ and \rosat\ data using XSPEC\footnote{http://xspec.gfsc.nasa.gov}~v12.7.
The quiescent flux of \src\ was determined by first extracting the source spectrum from the
\emph{ROSAT} data, then fitting it with an absorbed blackbody model. The
absorption column density $N_{\rm H}$ was fixed during the fit at the
best-fit value ($4.53\times 10^{21}\, \textrm{cm}^{-2}$) determined from the \swift\ and \cxo\ spectra (see below). 
We obtained a quiescent blackbody temperature of $kT=0.12\pm0.02$\,keV and a radius
of $5_{-2}^{+7}d_{1.6}$\,km, where $d_{1.6}$ is the distance to the source in units of 1.6\,kpc, 
the estimated distance as discussed in 
Section \ref{sec:distance}. We found an absorbed flux
of $9_{-9}^{+20}\times10^{-14}$\,erg\,cm$^{-2}$\,s$^{-1}$ in the
0.1--2.4\,keV range.

The \cxo\ and \swift\ spectra were grouped 
with a minimum of 100 and 20 counts per bin, respectively. 
The spectra were fitted jointly to a photoelectrically absorbed blackbody model with an added power-law component.
The model was fit with a single $N_{\rm H}$ using the XSPEC {\tt phabs} model assuming abundances from \citet{ag89} and
photoelectric cross-sections from \citet{bm92}. 
All the other parameters were allowed to vary from observation to observation.

The blackbody plus power-law model 
provided an acceptable fit 
to the \swift\ and \cxo\ data. The model had $\chi^2_\nu/\nu$ of 1.07/5451 and a best-fit $N_{\rm H}$ 
of $4.53\pm0.08\times 10^{21}\, \textrm{cm}^{-2} $.
Figure \ref{fig:spec} shows the evolution of the post-outburst spectral parameters.
The spectrum is softening following the outburst as the 1--10 keV absorbed flux 
(top panel of Figure \ref{fig:flux}) decays: the photon index of the power-law component
increases and the temperature of the blackbody decreases. This is clear from the high-quality
\cxo\ data alone but is also apparent in the \swift\ data, and is consistent with the 
behavior of other magnetars post-outburst (see Section \ref{sec:fluxdisc})

Figure \ref{fig:flux} shows the flux decay and pulsed fraction evolution following the \src\ outburst.
In \lsk, we showed that both a double exponential and an exponential model 
provided acceptable fits to both the total and pulsed flux decays, whereas a power-law decay model
was excluded. Here we fitted a double-exponential decay and an exponential decay
to the total and pulsed flux evolutions.
The exponential decay model is described
by $F(t) = F_p \exp^{-(t-t_0)/\tau} + F_q$ where $t$ is in MJD, $F_p$ is the peak absorbed flux,
$F_q$ is the quiescent flux, 
$t_0$ is the time of the BAT trigger in MJD, and $\tau$ is decay timescale in days.
The double-exponential decay is described by $F(t) = F_1 \exp^{-(t-t_0)/\tau_1} + F_2 \exp^{-(t-t_0)/\tau_2}+F_q$ 
where $t_0$ is the time of the BAT trigger in MJD, $F_1$ and $F_2$ are the absorbed fluxes at $t_0$ of each exponential 
component and $\tau_1$ and $\tau_2$ are the decay timescales in days of each component.
For both models we fit the data both by using the quiescent flux as a free parameter, and by using a fixed 
quiescent flux $F_q = 3\times10^{-14}$\,erg\,cm$^{-2}$\,s$^{-1}$.
This is the approximate 1--10\,keV flux assuming the 0.1--2.4\,keV flux and spectral model from the \rosat\ observation.  
Table \ref{ta:fluxmodel} shows the results of these fits. 
With $\chi_\nu^2/\nu$ of 1.06/40, the double-exponential decay with a free quiescent flux provides the best fit to
the data. However, the best-fit value for the quiescent flux, $4.0\pm0.2\times10^{-12}$\,erg\,cm$^{-2}$\,s$^{-1}$,
is more than two orders of magnitude higher than the quiescent flux implied by the \rosat\ observation. We therefore
take the double-exponential decay with the fixed quiescent flux as the best model of the total absorbed flux decay
with timescales of $\tau_1 = 15.5\pm0.5$ days and $\tau_2 = 177\pm14$ days. In both cases, 
the single-exponential fit is much worse than the double exponential.

In Figure \ref{fig:flux}, the absorbed flux measured with \cxo\ is always
larger than that with \swift\ at a similar epoch by about 10--15\%. The discrepancy could
be attributed to insufficient cross-calibration between instruments. \citet{tgp+11}
 found that flux measurements from different X-ray telescopes could
differ by as much as 20\%, and that \cxo\ appears to give a higher flux
than others, as well as a harder photon-index, which is consistent with our findings.  

To determine the 2--10 keV pulsed count rate from \swift, \rxte, and \cxo\ observations, the barycentered events 
were folded using the Solution 2 ephemeris in Table \ref{ta:coherent} with 16 phase bins. For the \rxte\ 
observations, only data from the first xenon layer of PCU2 were used. Both PCU0 and PCU1 lost their 
propane layers and there is minimal PCU3 and PCU4 data for this source. The inclusion of only a
single detector in the analysis reduces instrumental biases. 
For observations from all three telescopes,
the pulsed count rate was then measured from each folded profile using a RMS method as described in
\citet{dkg08}, where the pulsed count rate, $F$, is given by:
\begin{equation}
F = \sqrt{2 \sum^n_{k=1} [(a_k^2 + b_k^2) - (\sigma^2_{a_k} + \sigma^2_{b_k})]}\,\textrm{,}
\end{equation}
where $a_k$ is the even Fourier component and is equal to $(1/N)\sum^N_{i=1}p_i\cos(2\pi ki/N)$, $\sigma_{a_k}$
is the uncertainty in $a_k$, $b_k$ is the odd Fourier component and is equal to $(1/N)\sum^N_{i=1}p_i\sin(2\pi ki/N)$, $\sigma_{b_k}$
is the uncertainty in $b_k$, $i$ is an index over phase bins, $N$ is the total number of phase bins,
$p_i$ is the count rate in the $i$th phase bin, and $n$ is the maximum number of Fourier harmonics 
used. In this case, $n=5$. This technique is equivalent to the simple RMS formula,
$F=(1/\sqrt{N})[\sum^N_{i=1}(p_i-\bar{p})^2]^{1/2}$, except only statistically significant Fourier components
are used and the upward statistical bias is removed by subtracting the variances \citep{dkg08}.
For \swift\ and \cxo\ observations, pulsed fractions were 
determined by dividing the pulsed count rate by the total count rate.

The middle panel of Figure \ref{fig:flux} shows the 2--10\,keV pulsed-flux evolution of \linebreak\src.
The pulsed count rates measured by each instrument depend on the different instrumental responses.
The \rxte\ PCA and \cxo\ pulsed count rates were therefore scaled to the \swift\ WT mode values by 
including factors between each data set as free parameters in the double and 
single-exponential fits. 
For the pulsed-flux evolution, the double-exponential fit also provided
the best fit with $\chi_\nu^2/\nu$ of 5.15/74 and decay timescales of $\tau_1 = 15.3\pm0.2$ days and 
$\tau_2 = 182\pm6$ days. The exponential model had a $\chi_\nu^2/\nu$ of 60.5/76
with a best-fit decay timescale of $25.1\pm0.2$ days. This is the opposite of what we found in \lsk\
 where the exponential model was a better fit to the data available at the time.

\subsection{X-ray Bursts}
\label{sec:bursts}

To search for X-ray bursts in \rxte\ data of  \src,
we created a time series for each active PCU from {\tt{GoodXenon}}
data for each observation, selecting events in the
2~--~20\,keV range (PCA channels 6--24) and from all three detection layers
\citep[the same energy range as selected for similar searches for
X-ray bursts from magnetars, e.g.][]{gkw+01, gkw04}.
For the \swift\ observations, binned time series were made for each Good 
Timing Interval (GTI) in an observation.
For both \swift\ and \rxte, time series were made at
15.625-ms, 31.25-ms, 62.5-ms and 125-ms time resolutions to provide
sensitivity to bursts on a hierarchy
of time scales.

Bursts were identified by comparing the count rate
in the i$^{th}$ bin to the average count rate as described in \citet{gkw04}. 
Because the background
rate of the PCA typically varies over a single observation, we
calculated a local mean around the i$^{th}$ bin for \rxte. For \swift\
data, a mean was calculated for each GTI. We then
compared the count rate in the i$^{th}$ bin to the mean.
If the count rate in a single bin was larger than the local/GTI
average, the probability of such a count rate occuring by chance was
calculated. For \rxte\ data, the probability of the count rate in the
corresponding bin in the other active PCUs was also calculated
(whether or not the count rate in that bin was greater than the local
average). If a PCU was off during the bin of interest, its probability
was set to 1. We then found the total probability that a burst was
observed, by multipying the probabilities for each PCU together. 
If the total probability of an event was $P_{i,{\rm{tot}}} \le
0.01/N$ (where $N$ is the total number of time bins searched), it was 
flagged as a burst.

We found six bursts in \rxte\ data of \src. 
The burst properties are summarized in Table~\ref{ta:bursts}.
In the Table are the MJDs of each burst, the number of counts in a 
31.25-ms bin, and the probability that the burst would occur by chance
given the local mean count rate.
An insufficient number of bursts was detected
to perform a detailed statistical analysis of the burst properties for
\src. The bursts found were very narrow, typically only one or two
 31.25-ms bins wide, and not very fluent compared to typical magnetar
bursts \citep[see][]{gkw+01,gkw04,sk11,lkg+11}. No significant changes
in the long-term flux decay were observed at the times of these bursts.

Although in certain \swift\ observations we detected several bursts, 
these had much softer spectra than typical magnetar bursts and were also
seen in the background region. Therefore, we do not believe they originate 
from \src. No other bursts were detected in any of the \swift\ data.

\begin{deluxetable}{lccc}
\tablecaption{X-ray Bursts from \src.  \label{ta:bursts}}
\tablewidth{0pt}
\tabletypesize{\scriptsize}
\tablehead{\colhead{\rxte\ Obsid}&\colhead{MJD}&\colhead{Total
counts}& \colhead{Chance Prob$^a$}}
\startdata
\cutinhead{\rxte\ bursts}
D96048-02-01-01 & 55761.53224  & $15\pm4$ & $7.8\times10^{-7}$ \\
D96048-02-01-01 & 55761.57082  & $36\pm6$ & $8.6\times10^{-33}$ \\
D96048-02-01-02 & 55762.49919  & $21\pm5$ & $1.1\times10^{-13}$ \\
D96048-02-03-04 & 55777.91627  & $12\pm3$ & $4.5\times10^{-5}$  \\
D96048-02-04-01 & 55782.53122  & $13\pm4$ & $2.4\times10^{-5}$  \\
D96048-02-05-01 & 55789.96209  & $11\pm3$ & $2.2\times10^{-4}$  \\
\enddata
\tablecomments{$^a$The probability of the detected signal being due to noise. }
\end{deluxetable}

\section{Discussion}

We have presented \swift, \rxte, \cxo\ observations following the 
discovery of 
\linebreak\src\ during its outburst in 2011~July. 
We presented a phase-connected timing
solution which suggests a spin-down
inferred $B\sim5\times10^{13}$\,G, the second lowest measured for a magnetar thus far,
although we note that timing noise may significantly contaminate this estimate. 
The flux of the magnetar was found to be decaying, both in total and pulsed flux,
according to a double-exponential model. The spectrum softened
following the outburst. \src\ also emitted several short bursts 
during its period of outburst.
We also analysed
an archival \rosat\ observation from which \citet{erit11} previously reported
that \src\ is detected in quiescence. We note that the source had a similar absorption
column density to the nearby 
Galactic H{\sc ii} region M17. 
In the following we discuss the above results.

\subsection{Timing Behavior}

In Section \ref{sec:timing} we presented a timing solution for \src\ with
just $\nu$ and $\dot\nu$ fitted (Solution 1).
However, this solution appears significantly contaminated by timing noise,
a common phenomenon in pulsars. 
Most pulsars seem to display some unexplained
`wandering' in their spin evolution \citep{hlk10}. 
A measure of the amount of timing noise displayed by a pulsar is $\Delta_8$ and is defined as
$\Delta_8=\log[(1/6\nu)|\ddot{\nu}|(10^8{\mathrm s})^3]$ \citep{antt94}. \citet{hlk10} measured 
a correlation between $\dot{P}$ and $\Delta_8$ using timing solutions for 366 rotation-powered pulsars.
Magnetars are very noisy timers, generally having more timing noise, as measured by $\Delta_8$, than those 
rotation-powered pulsars of similar properties \citep{gk02,wkg+02}.
Here, for \src, we measure $\Delta_8=2.8$ (using $\ddot{\nu}$ from Solution 3)
which is much higher than the value
predicted from the correlation in \citet{hlk10} of $\sim-2$.
However, we caution that in general the $\ddot\nu$ used to calculate $\Delta_8$
is measured for a data span of $10^8$\,s, whereas our data span in much shorter.
The large value of $\Delta_8$ we measured may be biased by the short span,
or by unmodelled relaxation following a hypothetical glitch that could
have occured at the BAT trigger. Glitches are commonly seen to accompany radiative
outbursts from magnetars \citep[e.g.][]{kgw+03,dkg09}.

Due to the presence of timing noise, we take the timing and derived parameters of Solution 3 
not as the `true' spin-inferred
values, but as a `best guess' given the data thus far. As such, the uncertainties
in the parameters presented, which do not take into account the effect of contamination by timing noise,
likely underestimate the true uncertainties.
Further timing observations
of \src\ will help to average over the effects of timing noise
and thus provide improved estimates of the spin-inferred magnetic field of the pulsar.

The $B$-field measured by Solution 1 would be the second lowest measured for a magnetar to date, higher
than only SGR~0418+5729 \citep{ret+10}. Solution 3, although still the second lowest yet measured, gives a 
higher value of $B$ that is close to that of magnetar 1E~2259+586 and the magnetically active 
rotation-powered pulsar PSR~J1846$-$0258. It is also similar to the quantum critical field of
$B_{\mathrm QED}= 4.4\times10^{13}$\,G \citep{td96a} which has been viewed in the past of being a lower 
limit on the magnetic field of magnetars, although SGR~0418+5729 has shown that it is not a necessary condition
for magnetar-like activity.

\subsection{Flux and Spectral Evolution}
\label{sec:fluxdisc}

\begin{figure}
\plotone{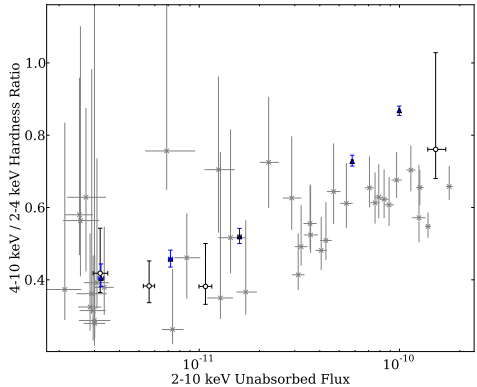}
\figcaption{
Spectral hardness as a function of unabsorbed flux for \src.
The \swift\ WT mode observations are shown as grey crosses, the
\swift\ PC mode data are denoted by open circles, and the blue
triangles respresent the \cxo\ observations.
\label{fig:hardening}
}
\end{figure}

In the twisted-magnetosphere model of magnetars, the thermal emission
is thought to originate from heating within the star, caused
by the decay of strong internal magnetic fields \citep{tlk02}. Currents in the magnetosphere,
which are due to twists in the magnetic field \citep{tlk02, bel09}, scatter the thermal
surface photons to higher energies. In addition to scattering, the currents provide
a source of surface heating in the form of a return current. The flux increase that accompanies a magnetar outburst
is theorized to be due a rapid heating which could originate from magnetospheric, internal, or crustal 
reconfiguration of the neutron star. This release may result in 
a significant increase in the surface temperature, in the return-current heating,
and in the twisting of the magnetic field. Thus, an increase in flux due to an internal
heat release should result in an increase of the hardness of the emission. This hardness-flux
correlation is in agreement with observations of several magnetars \citep{gh07,tgd+08,zkd+08,sk11}. 

\citet{sk11} explored the hardness-flux correlation for magnetar outbursts
by comparing the relation between fractional increase in 4--10/2--4\,keV hardness
ratio and fractional increase in 2--10\,keV unabsorbed flux for six different 
outbursts in four different magnetars. We present a similar plot here in 
Figure \ref{fig:hardening} for \src. Here, however, the hardnesses and fluxes are absolute
quantities and not fractional increases over quiescent values as in \citet{sk11},
as there is no quiescent observation of \src\ with the appropriate spectral
coverage. 
Figure \ref{fig:hardening} shows that \src\ softens as the flux decreases following
the outburst and so is in broad agreement with the hardness-flux correlation 
observed in other magnetar outbursts. 
This spectral softening with flux decline is clear also in Figure \ref{fig:spec},
where $kT$ declines and the power-law index $\Gamma$ increases as the flux drops.

\subsubsection{Magnetars in quiescence}

The quiescent flux of \src\ measured by \rosat\ in 1993 is about 3 orders of magnitude
lower than the peak flux measured following the outburst. 
Such large flux variations have been observed in several other magnetars
(e.g. 1E~1547$-$5408, XTE~J1810$-$197, AX~J1845$-$0258; \citealt{ims+04,ghbb04,tkgg06,sk11,bis+11}). 
Other magnetars, such as 1E~1841$-$045 \citep{zk10,lkg+11}, 4U~0142+61 \citep{gdk+10}, 
and 1RXS~J170849.0$-$400910 \citep{dkh08} have not exhibited
large flux variations, but are much brighter in quiescence than are the magnetars with
large outbursts. The cause of this difference is unclear.
\citet{pr12} suggest that there is a maximum luminosity that can be reached by a
magnetar during an outburst due to neutrino cooling dominating at high crust temperatures.
This helps to explain the differences in outburst magnitudes, but does not address
the wide range of quiescent luminosities.

Case in point, the magnetar 1E~2259+586 has spin properties that are likely quite similar to those of \src\,
but has a much higher quiescent luminosity. The magnetic field 
measured from spin-down for 1E~2259+586 is $5.9 \times 10^{13}$\,G \citep{gk02}, close to 
$B = 5.1\times10^{13}$\,G for \src\ as estimated by our Solution 3. 1E~2259+586 also went into
a period of outburst on 2002 June 18 where the flux increased by a factor of \gapp~20 \citep{wkt+04}. 
However, in quiescence, 1E~2259+586 is much brighter than \src\ with a quiescent 2--10\,keV luminosity
of $~2 \times 10^{34}$\,erg\,\,s$^{-1}$ \citep{zkd+08} compared to \lapp~$10^{31}$\,erg\,\,s$^{-1}$
for \src. 

One possibility is that the `true' magnetic fields of the more luminous magnetars are higher
than those of the fainter magnetars. 
The spin-down of the neutron star is only sensitive to the dipole
component of the magnetic field. If the magnetic field had significant components in higher
multipoles or a toroidal component \citep{td96a,pp11a}, the true magnetic field could be higher. 

Another possibility is that neutrino cooling in the core is setting a long-term luminosity limit, 
and that the neutrino cooling properties of the stars are different, 
e.g.~ due to different masses.
For example, consider first the case where the neutrino emission in the core is due to 
the modified URCA process,
with an emissivity $\epsilon_\nu\sim 10^{20}\ {\rm erg\ cm^{-3}\ s^{-1}}\  T_9^8$ \citep{ylh03}.
If we take the magnetic-field decay time to be $\tau=10^4\ {\rm yrs}$, then the luminosity from 
magnetic field decay is roughly $L_B=(B^2/8\pi)(4\pi R^3/3)(1/\tau)=10^{34}\ {\rm erg\ s^{-1}}$ 
for $B=10^{14}\ {\rm G}$. Balancing this with the neutrino losses $L_\nu=(4\pi R^3/3)\epsilon_\nu$,
we find a core temperature $T_c=2.5\times 10^8\ {\rm K}$ or, using the core temperature-luminosity 
relation from \citet{py01}, a luminosity $L\approx 4\times 10^{33}\ {\rm erg\ s^{-1}}$. 
On the other hand, if the neutrino emission is by the direct URCA process, with 
$\epsilon_\nu\sim 10^{26}\ {\rm erg\ cm^{-3}\ s^{-1}}\ T_9^6$ \citep{ylh03}, 
we find a core temperature $T_c=1.5\times 10^7\ {\rm K}$, corresponding to a surface luminosity 
of $\approx 2\times 10^{31}\ {\rm erg\ s^{-1}}$. 
This shows that we might reasonably expect a factor of $\gtrsim 200$ in luminosity 
between different stars if one has slow neutrino emission in the core, and the other fast, 
for example if the mass of one of the stars is large enough for direct URCA reactions 
to occur in the core. Even in the case where external currents dominate the quiescent luminosity, 
thermal emission from the neutron star provides a baseline luminosity, 
so that the low quiescent luminosity of \src\ suggests a low core temperature 
which implies either a low heating rate or efficient neutrino emission.

\subsubsection{The observed luminosity decay of \src}

\begin{figure}
\plotone{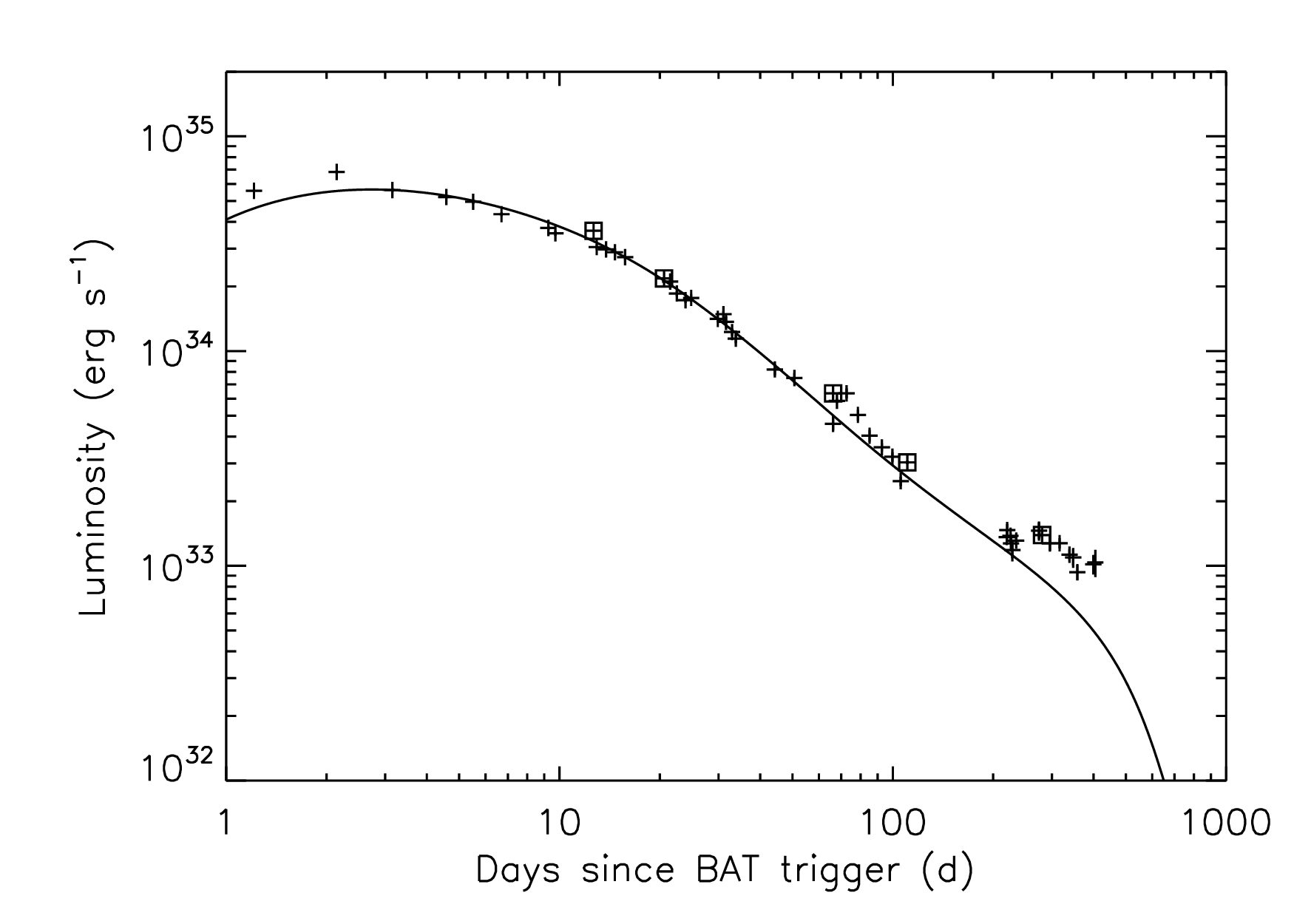}
\figcaption{
A model of the thermal relaxation of the neutron-star crust 
that approximately reproduces the observed 1--10 keV luminosity decay assuming 
a distance of 1.6\,kpc. In the model, we deposit an energy 
$3\times 10^{42}\ {\rm ergs}$ in the outer crust between densities 
of $2\times 10^9$--$3\times 10^{10}\ {\rm g\ cm^{-3}}$. 
The subsequent cooling of the crust gives a good match to the 
observed light curve.
\label{fig:fluxmodel}
}
\end{figure}

We find that the observed luminosity decay is well reproduced by models of thermal relaxation 
of the neutron-star crust following the outburst. An example is shown in Figure \ref{fig:fluxmodel}, 
which shows the cooling of the crust after an injection of $\approx 3\times 10^{42}\ {\rm ergs}$ of 
energy at low density $\approx 10^{10}\ {\rm g\ cm^{-3}}$ in the outer crust at the start of the outburst.
We follow the evolution of the crust temperature profile by integrating the thermal diffusion equation. 
The calculation and microphysics follow \citet{bc09} who studied transiently accreting neutron stars, 
but with the effects of strong magnetic fields on the thermal conductivity included \citep{pbhy99} and 
for the outer boundary condition using the $T_{\rm eff}$--$T_{\rm int}$ 
relation appropriate for a magnetized envelope following \citet{py01}.
The calculation follows the radial structure only; we assume that the magnetic field geometry is a 
dipole and take appropriate spherical averages to account for the variation in thermal conductivity 
across the star \citep{pbhy99}. We assume $B=6\times 10^{13}\ {\rm G}$, similar to the value inferred 
from the spin down, a 1.6 $M_\odot$, $R=11.2\ {\rm km}$ neutron star, and take an impurity parameter 
for the inner crust of $Q_{\rm imp}=10$  \citep{pj04}. We set the neutron-star core temperature to 
$2\times 10^7\ {\rm K}$, which is needed to obtain a quiescent luminosity $<10^{32}\ {\rm erg\ s^{-1}}$.

With the neutron-star parameters fixed, we then vary the location and strength of the heating and find
that we obtain good agreement with the observed light curve for times $<100$ days, if the initial 
heating event is located at low densities $\lesssim 3\times 10^{10}\ {\rm g\ cm^{-3}}$. This conclusion 
comes from matching the observed timescale of the decay, and is not very sensitive to the choice of 
neutron-star parameters. For example, changing the neutron-star gravity changes the crust thickness 
and therefore cooling time, giving an inferred maximum density $\rho_{\rm max}\propto g^{-2}$. 
This means that the inferred density can change by a factor of a few but cannot be moved into the 
neutron drip region, for instance. That only a shallow part of the outer crust is heated is an 
interesting constraint on models of crust heating in a magnetar outburst.  
We find that it is difficult to match the observed light curve at times $\gtrsim 200$ days, 
but the late time behaviour of the light curve is sensitive to a number of physics inputs 
associated with the inner crust, including the thermal conductivity and superfluid parameters, 
as well as modification due to the angular distribution of the heating over the surface of the star. 
We will investigate the late-time behaviour in more detail in future work. Figure \ref{fig:fluxmodel} 
suggests that the source could undergo significant further cooling in the coming years.

\subsection{Distance Estimate and Possible Association}
\label{sec:distance}

As shown in the \rosat\ image (Figure \ref{fig:rosat}), the Galactic H{\sc ii} region
M17 is located $\sim$20\arcmin\ southwest of \src. It has a distance of
$1.6\pm0.3$\,kpc \citep{ncjm01} and an absorption column density
$N_{\rm H}=4\pm1\times10^{21}\,$cm\,$^{-2}$ \citep{tfm+03} which
is consistent with our best-fit value of $4.53\pm0.08\times10^{21}$\,cm\,$^{-2}$.
This suggests that
\src\ could have a comparable distance to that of M17\footnote{While there are two
molecular clouds surrounding M17 \citep{whm03}, they are confined to
the north and west, such that they should not contribute to the $N_{\mathrm H}$ of either
M17 or \src.}. If so, then \src\ would be
one of the closest magnetars detected thus far. 

The above argument does not necessitate a direct association between M17 and \src.
However, if \src\ is associated with M17, then its angular separation of 26\arcmin\
from the cluster center, where the X-ray emission peaks in the \emph{ROSAT}
image, implies a physical distance of 12\,pc. For a pulsar age of $10^5$\,yr,
this requires a space velocity of only $\sim$100\,km\,s$^{-1}$ (corresponding
to a proper motion of 0\farcs016\,yr$^{-1}$).
This would make a direct proper motion measurement difficult. 
From timing, the characteristic age appears to be larger than $10^5$\,yr which
would further reduce the implied proper motion. On the other hand, characteristic
ages can be large overestimates of the true age. However, even if the true age
were as low as $10^4$\,yr, the proper motion would be difficult to measure
even with \chandra. Additionally, if the magnetar was born near an edge of
the cluster, the angular separation from its birthplace could be larger or
smaller by up to $\sim10$\arcmin.

\section{Conclusions}

We have presented an analysis of the post-outburst radiative evolution and timing behavior of \src,
following its discovery on 2011 July 14. 
Following a timing analysis for the source post-outburst, we estimate the surface dipolar
component of the $B$-field to be $\sim5 \times 10^{13}$\,G, slightly higher than that inferred
in \lsk.
However, as this measurement is contaminated by timing noise, the true
value of the magnetic field could be well outside of the uncertainties quoted in Table \ref{ta:coherent}. 
Futher monitoring of \src\ as it fades following the outburst
will allow us to better account for the timing noise and measure more robust
timing parameters. 

The quiescent flux of \src\ measured using a 1993 \rosat\ observation of M17 was
found to be roughly three orders of magnitude lower than the peak flux during the outburst. 
The flux evolution following the outburst was 
well characterized by a double-exponential
decay. By applying a crustal cooling
model to the flux decay, we found that the energy deposition likely occured in the outer crust
at a density of $\sim10^{10}$\,g\,cm$^{-3}$.
The spectral properties of \src\ were observed to soften following the outburst, with the
power-law index increasing and the temperature of the blackbody decreasing.
Indeed, a hardness-flux correlation, similar to what is observed in other magnetars, was
clearly observed.
Based on the similarity in $N_H$ to that of the H{\sc ii} region M17, we argue
for a source distance of $1.6\pm0.3$ kpc, one of the closest distances yet inferred for
a magnetar.\\

We are grateful to the \swift, \chandra, and \rxte\ teams for their flexibility
in scheduling TOO observations. We thank the anonymous referee for helpful comments and suggestions.
V.M.K. holds the Lorne Trottier Chair in Astrophysics and Cosmology and a Canadian
Research Chair in Observational Astrophysics. This work is supported by NSERC via a Discovery Grant, by FQRNT
via the Centre de Recherche Astrophysique du Qu\'ebec, by CIFAR,
and a Killam Research Fellowship.

\bibliographystyle{apj}

\clearpage

\end{document}